\documentclass[letterpaper, paper,11pt]{AAS}	

\usepackage{bm}
\usepackage{amsmath}
\usepackage{subfigure}
\usepackage[colorlinks=true, pdfstartview=FitV, linkcolor=black, citecolor= black, urlcolor= black]{hyperref}
\usepackage{overcite}
\usepackage{footnpag}			      	

\usepackage{wasysym}
\usepackage{siunitx}

\NewDocumentCommand{\mathleftmoon}{}{{\text{\normalfont\leftmoon}}}

\usepackage{graphicx}
\usepackage{amsfonts,mathtools}
\usepackage{amsmath}
\usepackage{amsthm}

\usepackage{siunitx}
\usepackage{longtable,tabularx}
\usepackage{cleveref}

\DeclareMathOperator*{\argmin}{arg\,min}

\def\*#1{\boldsymbol{\mathbf{#1}}}
\def\##1{\bm{\mathsf{#1}}}

\usepackage{tikz}
\usetikzlibrary{calc}
\usetikzlibrary{math}

\newcommand{\appropto}{\mathrel{\vcenter{
  \offinterlineskip\halign{\hfil$##$\cr
    \propto\cr\noalign{\kern2pt}\sim\cr\noalign{\kern-2pt}}}}}

\DeclarePairedDelimiterX{\norm}[1]{\lVert}{\rVert}{#1}

\usepackage{pgfplots}
\pgfplotsset{compat=1.18} 
\usepackage{pgfplotstable}
\usepackage{cbcolors}

\usepackage{newunicodechar,graphicx}

\DeclareRobustCommand{\okina}{%
  \raisebox{\dimexpr\fontcharht\font`A-\height}{%
    \scalebox{0.8}{`}%
  }%
}
\newunicodechar{ʻ}{\okina}

\begin{document}

\pgfplotsset{clean/.style={axis lines*=left,
        axis on top=true,
        axis x line shift=0.0em,
        axis y line shift=0.75em,
        every tick/.style={black, thick},
        axis line style = ultra thick,
        tick align=outside,
        clip=false,
        major tick length=4pt}}

\title{Deterministic Optimal Transport-based Gaussian Mixture Particle Filtering for Verifiable Applications}

\author{Andrey A. Popov\thanks{Assistant Professor, Department of Information and Computer Sciences, The University of Hawai\okina i at M\=anoa, Honolulu, HI 96822.},  
\ and Renato Zanetti\thanks{Associate Professor, Department of Aerospace Engineering and Engineering Mechanics, The University of Texas at Austin, Austin, TX 78712.}
}

\maketitle

\begin{abstract}
Mixture-model particle filters such as the ensemble Gaussian mixture filter require a resampling procedure in order to converge to exact Bayesian inference. 
Canonically, stochastic resampling is performed, which provides useful samples with no guarantee of usefulness for a finite ensemble.
We propose a new resampling procedure based on optimal transport that deterministically selects optimal resampling points. 
We show on a toy 3-variable problem that it significantly reduces the amount of particles required for useful state estimation. 
Finally, we show that this filter improves the state estimation of a seldomly-observed space object in an NRHO around the moon. 
\end{abstract}

\section{Introduction}

The invocation of the term `probability' conjures up a sense that some type of `random' process is naturally involved. This is not necessarily the case. It is perfectly possible to deal with probability, even complex probability, in a fully deterministic fashion. This work shows that it is even possible to build a fully convergent, deterministic,  particle filter that requires only a few particles to converge to exact Bayesian inference.

The ensemble Gaussian Mixture filter (EnGMF) is a particle filter that combines the ideas of Gaussian mixtures and particle filtering to create a low-cost convergent particle filter that requires significantly less particles than the bootstrap particle filter (BPF).~\cite{popov2023elengmf2,popov2024adaptive,yun2022kernel}
Just like the BPF, the EnGMF relies on a resampling step in order to represent our knowledge through a collection of equally weighted particles.

A recent development in particle filtering is the optimal transport-based ensemble transform particle filter (ETPF).~\cite{reich2013nonparametric}
Instead of resampling, the ETPF transports the prior particles into ones that are equally weighted in the posterior. 
The advantage of this strategy is that the transport map is a deterministic linear transformation of the particles from the prior to the posterior.
The ETPF methodology is known to converge to exact Bayesian inference (in distribution).~\cite{reich2015probabilistic}
The limitation of the ETPF methodology is that is still requires the use of stochastic regularization techniques and requires a significant amount of particles to practically converge.

The stochastic shrinkage technique augments the ensemble of particles through random samples from known good (``climatological'') distributions in a statistically consistent manner.~\cite{popov2022stochastic}
The limitation of the stochastic shrinkage technique is that it requires sampling from a precomputed probability distribution, and only converges when the impact of this distribution goes to zero.

Combining the ideas of optimal transport particle filtering and the stochastic shrinkage technique, we aim to create a filter that is both able to transport particles in a deterministic fashion, but also take advantage of the Gaussian mixture structure of the EnGMF to sample from an online distribution, removing the limitations of the stochastic shrinkage approach.

The advantages of a deterministic approach to particle filtering are many-fold: \textit{i)} it becomes possible to verify that the particle filter will perform the same exact way, without relying on a (pseudo) random number generator, \textit{ii)} the error from resampling a finite number of particles is entirely taken out of consideration, and \textit{iii)} the particle filter becomes a much for realistic choice for mission critical applications.

\section{Background}

Assume that we have knowledge whose distribution is a Gaussian mixture model (GMM), 
\begin{equation}\label{eq:GMM}
    p^-(x) = \sum_{i=1}^N w_i^-\, \mathcal{N}(x; \mu_i^-, \Sigma_i^-),
\end{equation}
consisting of $N$ components with weights $\{w_i\}_{i=1}^N$, means $\{\mu_i\}_{i=1}^N$, and covariances $\{\Sigma_i\}_{i=1}^N$. Such a Gaussian mixture model can be obtained through kernel density estimation techniques from a given set of $N$ samples,
\begin{equation}
\begin{aligned}
    X &= [x_1, \dots, x_N],\\
    \mu_i^- &= x_i,\\
    \Sigma_i^- &= \beta^2 \operatorname{Cov}(X),\\
    w_i^- &= \frac{1}{N},
\end{aligned}
\end{equation}
for all $i = 1, \dots, N$, and where $ \beta^2$ is known as the bandwidth parameter and for this work is defined as,
\begin{equation}
     \beta^2 = {\left(\frac{4}{N(n+2)}\right)}^{\frac{2}{n+4}},
\end{equation}
taken from~\cite{silverman2018density}. This value of the bandwidth parameter is optimal when the underlying distribution of the samples is Gaussian, but still allows the density estimate to converge when the distribution is non-Gaussian, as $ \beta^2 \to 0$ as $N\to\infty$.

Typical sampling procedures attempt to find $L$ samples from~\cref{eq:GMM}.
The goal of this work is more restrictive: to find a collection of $N$ equally weighted samples---one for each component of the GMM~\cref{eq:GMM}---in a deterministic fashion, so that the pseudo-resampling procedure could be used in a mission critical environment.

\subsection{Gaussian sum update}
In the Gaussian sum update, each Gaussian component of~\cref{eq:GMM} is updated using some sort of Gaussian (two moment) filter, 
\begin{equation}
    \mathcal{N}(x ;\mu^+, \Sigma_i^+) \appropto \mathcal{N}(x ;\mu^-, \Sigma_i^-)\,\mathcal{N}(y; h(x), R),
\end{equation}
where $h$ is the non-linear measurement operator, $y$ is the measurement, and $R$ is the measurement covariance.
This is typically performed using the extended Kalman filter though a linearization of the measurement operator around the prior mean.
The weights are proportional to
\begin{equation}
    w^+_i \propto w^-_i \int_{\mathbb{R}^n}  \mathcal{N}(x ;\mu^-, \Sigma_i^-)\,\mathcal{N}(y; h(x), R), \mathrm{d} x,
\end{equation}
typically performed in a linearized way, mimicking the EKF.~\cite{durant2024you}
The weights can be regularized towards uniformity with a defensive factor~\cite{popov2024adaptive},
\begin{equation}\label{eq:defensive-factor}
    w^+_i \xleftarrow[]{} (1 - \mathfrak{v}_N)w^+_i + \mathfrak{v}_N/N
\end{equation}
where $\mathfrak{v}_N \to 0$ as $N\to\infty$, and is bounded $0 < \mathfrak{v}_N < 1$.

This results in the posterior GMM,
\begin{equation}\label{eq:posterior-GMM}
    p^+(x) = \sum_{i=1}^N w_i^+\, \mathcal{N}(x; \mu_i^+, \Sigma_i^+),
\end{equation}
which describes our posterior knowledge about the state of interest.
For a more detailed exposition on the EnGMF, the Gaussian sum update and a discussion of the weights see~\cite{popov2024adaptive, durant2024you,popov2024epanechnikov}.

\subsection{Stochastic Resampling}

The goal of stochastic resampling is to produce an ensemble, of $N$ independent and identically distributed samples from the posterior Gaussian mixture~\cref{eq:posterior-GMM}. 
The procedure for performing na\"ive stochastic resampling is quite simple,
\begin{enumerate}
    \item sample the Gaussian component from the discrete distribution defined by the posterior GMM weights, $\{w_i^+\}_{i=1}^N$,
    \item for each sampled component $\mathcal{N}(\mu_\ell, \Sigma_\ell)$ take a sample therefrom
    \item repeat until the desired number of samples ($N$) is produced.
\end{enumerate}

The resulting ensemble,
\begin{equation}
    X^+ = \left\{x^+_1, \dots, x^+_N\right\},
\end{equation}
defines an empirical measure,
\begin{equation}\label{eq:stochastic-posterior-empirical-measure-converge}
    \sum_{i=1}^N w^+_i\mathcal{N}(x ; \mu_i^+, \Sigma_i^+) \xleftarrow[]{\mathcal{D}} \sum_{i=1}^N  \frac{1}{N}\delta_{x_{i}^+}(x),
\end{equation}
that converges in distribution to the posterior GMM~\cref{eq:posterior-GMM}.

The advantage of the stochastic resampling procedure is that it is computationally simple and embarrassingly parallelizable.
The disadvantage of the stochastic resampling procedure is that there are no guarantees that the samples are a good representation of either the Gaussian mixture kernel density estimate, or the underlying distribution from which the estimate was derived. In the ensemble limit ($N\to\infty$) the stochastic resampling procedure is equivalent to bootstrap sampling, and is thus convergent to exact Bayesian inference.

These samples are them typically propagated to the next step, the Gaussian mixture kernel density estimate is rebuilt, the Gaussian sum update is performed and subsequently resampled and the process repeats. This sequence of steps is known as the Ensemble Gaussian Mixture filter (EnGMF).

\subsection{Optimal transport}

\begin{figure}[t]
    \centering
    \includegraphics[width=0.65\linewidth]{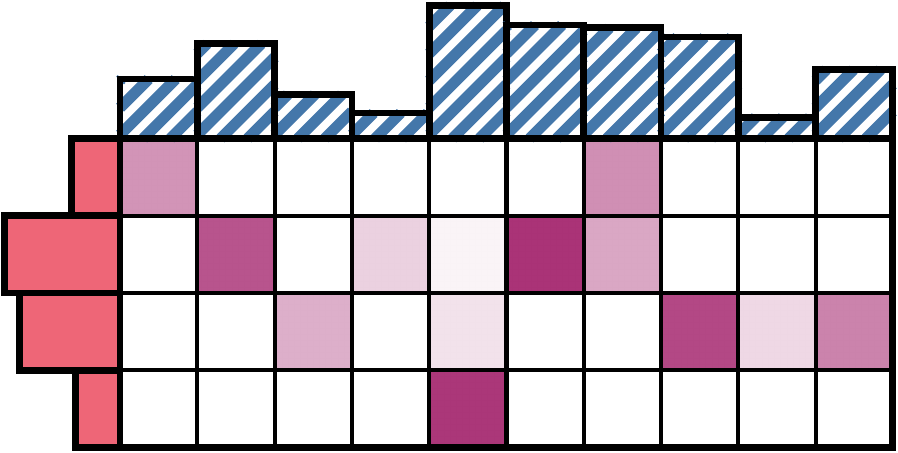}
    \caption{Illustration of the optimal transport framework. The blue top bars with diagonal stripes represent the discrete origin distribution, the red left bars represent the destination distribution, and the purple shaded squares in the middle represent the transport plan.}
    \label{fig:optimal-transport}
\end{figure}

We can define the transport of mass from one probability distribution to another by a linear transport plan. Given a set cost to transport between the elements of the supports, the optimal transport plan is one that minimizes the total cost.~\cite{monge1781memoire,kantorovich1942translocation,bogachev2012monge}
Formally, given an ensemble of $P$ particles, 
\begin{equation}
    \widetilde{X} = \left\{\widetilde{x}_1, \dots, \widetilde{x}_P\right\},
\end{equation}
with the corresponding weights $\{\widetilde{w}_i\}_{i=1}^P$ (summing to unity), the optimal linear transport plan is,
\begin{equation}
    \widehat{X} = \widetilde{X} T^*,\quad T^* \in \mathbb{zp}^{M\times N}
\end{equation}
where the $P$ samples of $\widetilde{X}$ are transformed into the $N$ samples $\widehat{X}$ by minimizing the cost,
\begin{equation}\label{eq:optimal-transport}
    \begin{gathered}
        T^* = \argmin_{T} \sum_{i,j} T_{i,j} \, \norm{\widetilde{X}_i - \widetilde{X}_j}_{M}^2,\\
        \text{subject to}\,\, T1 = \widetilde{w},\,\, T^\texttt{T} 1 = \widehat{w},
    \end{gathered}
\end{equation}
where the constraints ensure that the mass is transported into a distribution with target weights $\{\widehat{w}_i\}_{i=1}^N$. The optimal transport methodology is visually described in figure~\cref{fig:optimal-transport}.

The ensemble transform particle filter (ETPF) makes use of optimal transport as a resampling procedure for the sequential importance resampling (SIR) family of particle filters.~\cite{reich2013nonparametric}
Next, we extend this idea to the ensemble Gaussian mixture particle filter.

\section{The Deterministic Resampling Methodology}

\begin{figure}[t]
    \centering
    \includegraphics[width=0.99\linewidth]{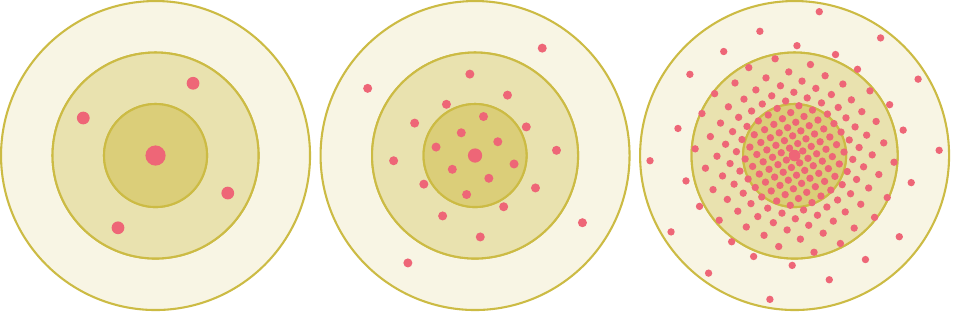}
    \caption{A visual representation of the Fibonacci grid Gaussian sampling procedure for $N=5$, $N=25$, and $N=201$ samples from left to right, respectively. The large point represents the mean of the standard 2-dimensional Gaussian (zero mean) and the other points represent deterministic samples therefrom. The shaded outlines represent one, two, and three standard deviations, ordered radially outwards from the mean.}
    \label{fig:Fibonacci-grid-example}
\end{figure}

The goal of the stochastic resampling procedure is to create samples that create an empirical measure~\cref{eq:stochastic-posterior-empirical-measure-converge} that converges to the posterior GMM~\cref{eq:posterior-GMM} in distribution. While it certainly achieves this goal, it is by no means the only way. In general, deterministic sampling approaches create empirical measures that also converge in distribution to the distribution of interest.~\cite{Diss24_Frisch}

We take the continuous posterior Gaussian mixture model~\cref{eq:posterior-GMM} and transform it into an approximate discrete representation. We make use of deterministic sampling techniques for Gaussian distributions, in particular the Fibonacci grids in order to sample from the constituent Gaussians~\cite{frisch2023generalized}.

As can be seen in~\cref{fig:Fibonacci-grid-example}, the  Fibonacci grids methodology creates a uniformly weighted, determinsitic grid of points, that (sub)-optimally constructs a regular approximation of a standard normal Gaussian. Through a trivial affine transformation, this grid can be made to represent a Gaussian with any mean and (full rank) covariance. 

We generalize this approach to Gaussian mixtures using a sampling of $M$ point for each component of the mixture, thus having $N\times M$ total points in the mixture.
For the $i$th component of the posterior GMM~\cref{eq:posterior-GMM} we construct a Fibonacci grid of $M$ points (where $M$ is taken to be odd such that the mean is one of the grid points),
\begin{equation}
    \{ x_{i,j}\}_{j=1}^M \sim_{\text{fib}} \mathcal{N}(x ; \mu_i^+, \Sigma_i^+),
\end{equation}
such that the component is well-approximated in distribution by the empirical measure resulting therefrom,
\begin{equation}
    \mathcal{N}(x ; \mu_i^+, \Sigma_i^+) \xleftarrow[]{\mathcal{D}} \sum_{j = 1}^M \frac{1}{M}\delta_{x_{i,j}}(x).
\end{equation}

The advantage of Fibonacci grids is their computational simplicity as they do not require the solution of any non-linear problem and merely require simple linear propagation to achieve.

The resulting collection of $M\times N$ points 
\begin{equation}\label{eq:full-deterministic-collection}
    \{x_{i,j}\}_{i=1,j=1}^{N,M},
\end{equation}
defines a deterministic empirical measure,
\begin{equation}
    \sum_{i=1}^N w^+_i\mathcal{N}(x ; \mu_i^+, \Sigma_i^+) \xleftarrow[]{\mathcal{D}} \sum_{i=1}^N \sum_{j = 1}^M \frac{w^+_i}{M}\delta_{x_{i,j}}(x),
\end{equation}
that converges (in $M$) to the posterior GMM~\cref{eq:posterior-GMM}.

We then transport the mass of the entire collection~\cref{eq:full-deterministic-collection} into the optimal $N$ points that describe the GMM~\cref{eq:posterior-GMM}. In order to accomplish that, we need to choose points that best represent the GMM from this large collection. As we chose $M$ to be odd, the means $\{\mu^+_i\}_{i=1}^N$ of the posterior GMM are members of the collection~\cref{eq:full-deterministic-collection} of deterministic samples from the GMM, thus they make the most obvious choice of points to transform.

By making use of~\cref{eq:optimal-transport}, we construct an optimal transport plan $T^*$ such that,
\begin{equation}
    X^+ = \operatorname{mat}\left(\{x_{i,j}\}_{i=1,j=1}^{N,M}\right)T^*,
\end{equation}
where the $\operatorname{mat}$ operator maps the collection onto a matrix with the first $N$ columns representing the means of the posterior GMM~\cref{eq:posterior-GMM}, and the subsequent columns representing the rest of the Fibonacci grid samples. Thus the columns of $X^+$ are the optimal $N$ points that represent, as deterministic samples, the posterior GMM~\cref{eq:posterior-GMM}.

\subsection{Particle Propagation and KDE}

Take the propagation equation,
\begin{equation}
    x^-_k = f\left( x^+_{k-1} \right) + \nu_k,
\end{equation}
where $\nu_k\sim\mathcal{N}(0,Q_k)$ is our representation of the model propagation uncertainty. 
In standard particle filter methodology, the propagation of each particle would take a unique sample $\nu_k$ for each particle.
In the deterministic sampling approach this is not desirable. 

Instead of adding a sample from the process noise distribution, we can add the noise to the kernel density estimate of the prior,
\begin{equation}
    \Sigma_{k}^- =  \beta^2 \operatorname{Cov}(X^-_k) + Q_k,
\end{equation}
thus avoiding any random number generation.

Propagating particles with non-additive process noise in an exact and deterministic manner is of separate, independent, interest.

\subsection{Illustration with a pineapple distribution}

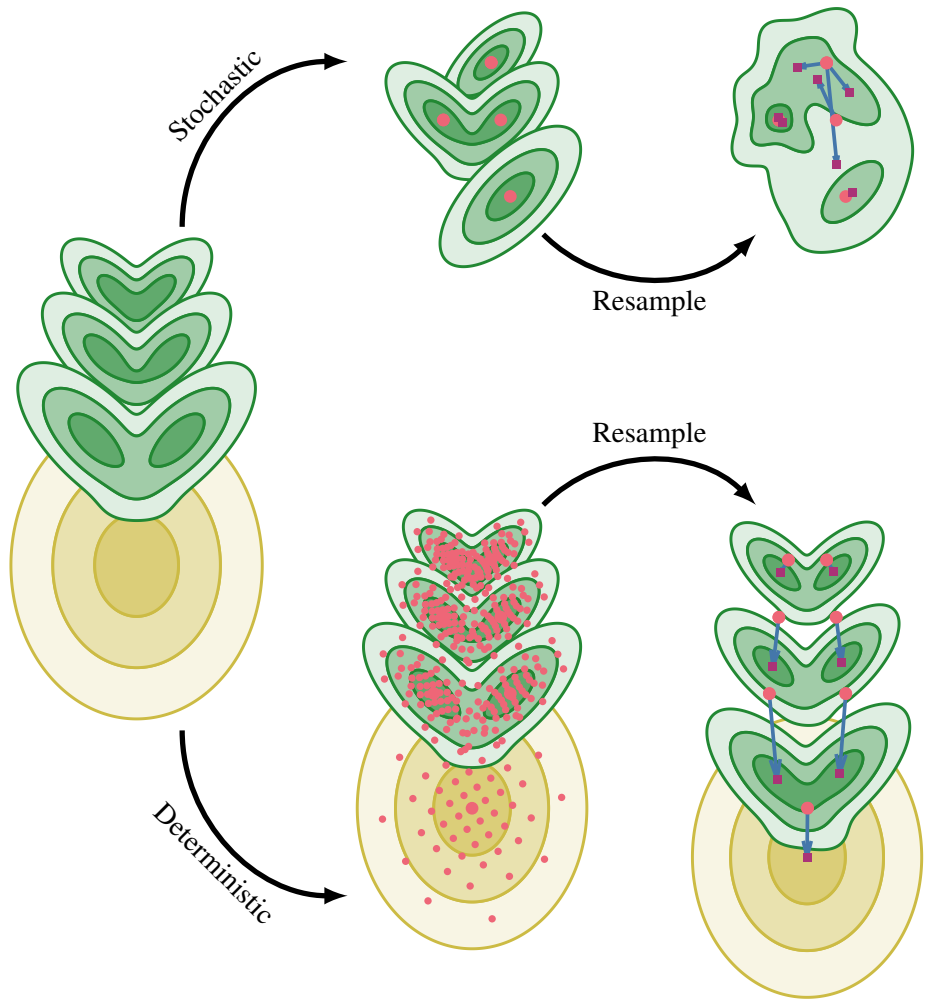
\begin{figure*}
    \centering
    \begin{tikzpicture}[scale=1, every node/.style={scale=1}]
        \node (pic) at (0,0) {\includegraphics[width=0.8\linewidth]{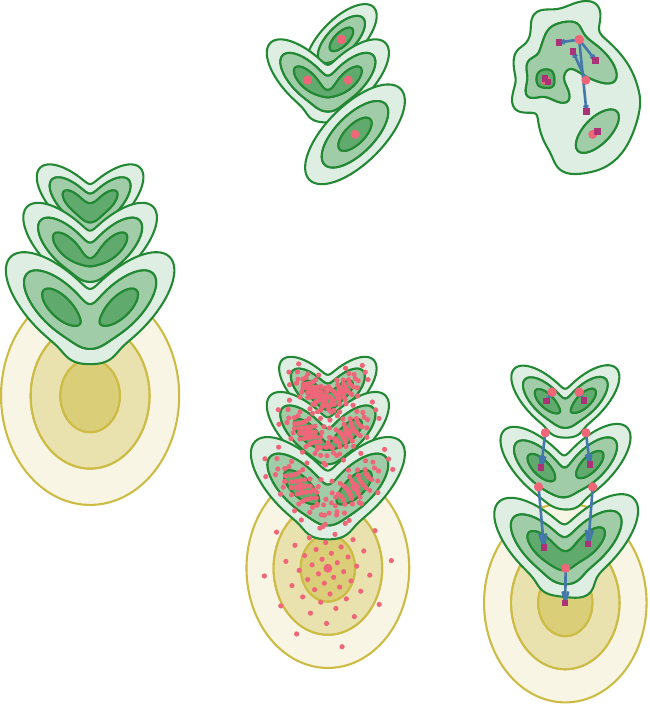}};
        \draw[black,-latex,line width=2pt] (-3.8,-3) arc (-180:-90:2.2) node[pos=.5,below,sloped] () {Deterministic};
        \draw[black,-latex,line width=2pt] (-3.8,3.7) arc (180:90:2.2) node[pos=.5,above,sloped] () {Stochastic};

        \draw[black,-latex,line width=2pt] (1,0) arc (135:45:2) node[pos=.5,above,sloped] () {Resample};

        \draw[black,-latex,line width=2pt] (1,3.6) arc (-135:-45:2) node[pos=.5,below,sloped] () {Resample};
    \end{tikzpicture}
    \caption{A representation of a Gaussian mixture distribution that looks like a pineapple. The left most figure represents the unweighted Gaussian mixture model from which we wish to resample, with the contour lines describing one, two, and three sigma bounds respectively. The top part of the figure represents the stochastic resampling procedure}
    \label{fig:pineapple-figure}
\end{figure*}

We now illustrate the deterministic resampling procedure with a fun example.
Consider the Gaussian mixture,
\begin{equation}
    p(x) = \sum_{i=1}^7 w_i\,\mathcal{N}(m_i, C_i),
\end{equation}
where the means,
\begin{equation}
\begin{gathered}
    m_1 = \begin{bmatrix}
            0\\ 0
        \end{bmatrix},\, m_2 = \begin{bmatrix}
            1\\ 3
        \end{bmatrix},\, m_3 = \begin{bmatrix}
            -1\\ 3
        \end{bmatrix},\, m_4 = \begin{bmatrix}
            0.75\\ 5
        \end{bmatrix},\\
     m_5 = \begin{bmatrix}
            -0.75\\ 5
    \end{bmatrix},\, m_6 = \begin{bmatrix}
            0.5\\ 6.5
        \end{bmatrix},\, m_7 = \begin{bmatrix}
            -0.5\\ 6.5
        \end{bmatrix},
\end{gathered}
\end{equation}
covariances, 
\begin{equation}
        C_1 =\begin{bmatrix}
            1 & 0\\ 0 & 1.5
        \end{bmatrix}, \quad C_i = 
        \frac{1}{\lfloor \frac{i}{2} \rfloor + 1}\begin{bmatrix}
            0.75 & (-1)^i 0.5\\ (-1)^i 0.5 & 0.75
        \end{bmatrix},\, i = 2,\dots, 7,
\end{equation}
and weights,
\begin{equation}
    w_i = \begin{cases}
        0.4 & i = 1\\
        0.1 & i = 2, \dots 7
    \end{cases},
\end{equation}
fully define what we call the pineapple distribution.

As the distribution consists of $N = 7$ components, the goal of a resampling procedure is to get the seven points that optimally represent the distribution. 
We compare the deterministic resampling approach with $M=51$ synthetic points per component to the stochastic resampling technique. 
The results of which are visually displayed in~\cref{fig:pineapple-figure}.
As can be seen, the stochastic resampling procedure, by necessity completely misses three out of the four modes, and produces samples that are a poor representation of the underlying distribution.
The deterministic resampling procedure, on the other hand, enforces one equally-weighted sample from each distribution and represents the underlying distribution in a recognizable manner.

As the example makes use of a distribution that looks like a pineapple, for lack of a better name, the author christen the EnGMF using deterministic resampling the `Pineapple Filter'.

\section{Sequential Filtering Experiments}

The numerical sequential filtering experiment aim to show two different results. First using the Lorenz '63 equations we show that the amount of samples requires for a convergent filter is significantly reduced compared to other particle filters.
Second, using a more realistic cislunar problem, we show that the deterministic sampling approach can handle a scenario for which the unscented Kalman filter diverges, and for which the ensemble Gaussian mixture filter (stochastic resampling) cannot converge for the given number of particles.

\subsection{Lorenz '63}

\begin{figure}[t]
    \centering
    \includegraphics[width=0.5\linewidth]{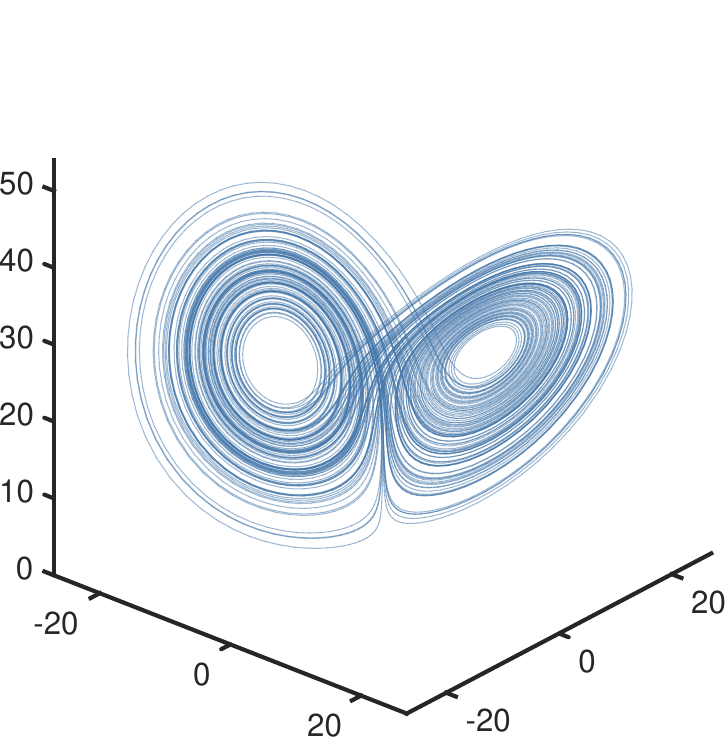}
    \caption{A visual representation of the three variable Lorenz '63 attractor.}
    \label{fig:lorenz-63-dynamics}
\end{figure}

\begin{figure}[t]
    \centering
    \begin{tikzpicture}[scale=0.85]
    \begin{axis}[clean,
        cycle list name=tol,
        xtick={4, 8, 12, 16, 20, 24, 30},
        table/col sep=comma,
        xmin = 1,
        xmax = 32,
        ymin = 0,
        ymax = 13.9,
        clip = true,
        xlabel = {Number of Particles (N)},
        ylabel = {Mean Spatio-temporal RMSE},
        every axis plot/.append style={line width=2pt, mark size=3.5pt},
        legend style={at={(0.5,0.88)},anchor=center},
        legend cell align={left}]

    \addplot[mark=o,color=tolyellow] table [x=Ns, y=mPineapple, col sep=comma] {data/Lorenz63experimentPineapple.csv};
    \addlegendentry{Pineapple Filter (Deterministic)}; 

    \addplot[mark=x,color=tolpurple] table [x=Ns, y=mEnGMF, col sep=comma] {data/Lorenz63experimentPineapple.csv};
    \addlegendentry{EnGMF (Stochastic)};

    \end{axis}
    \end{tikzpicture}%
    \begin{tikzpicture}[scale=0.85]
    \begin{axis}[clean,
        cycle list name=tol,
        xtick={4, 8, 12, 16, 20, 24, 30},
        table/col sep=comma,
        xmin = 1,
        xmax = 32,
        ymin = 0,
        ymax = 13.9,
        clip = true,
        xlabel = {Number of Particles (N)},
        ylabel = {Spatio-temporal RMSE 3-$\sigma$},
        every axis plot/.append style={line width=2pt, mark size=3.5pt},
        legend style={at={(0.5,0.88)},anchor=center},
        legend cell align={left}]

    \addplot[mark=o,color=tolyellow] table [x=Ns, y=s3Pineapple, col sep=comma] {data/Lorenz63experimentPineapple.csv};
    \addlegendentry{Pineapple Filter (Deterministic)}; 

    \addplot[mark=x,color=tolpurple] table [x=Ns, y=s3EnGMF, col sep=comma] {data/Lorenz63experimentPineapple.csv};
    \addlegendentry{EnGMF (Stochastic)};

    \end{axis}
    \end{tikzpicture}
    
    \caption{Number of particles ($N$) versus spatio-temporal RMSE for the Lorenz '63 problem. The left figure plots the mean of the RMSE across the 192 Monte Carlo trials, while the right figure plots three standard deviations.}
    \label{fig:lorenz63-experiment}
\end{figure}
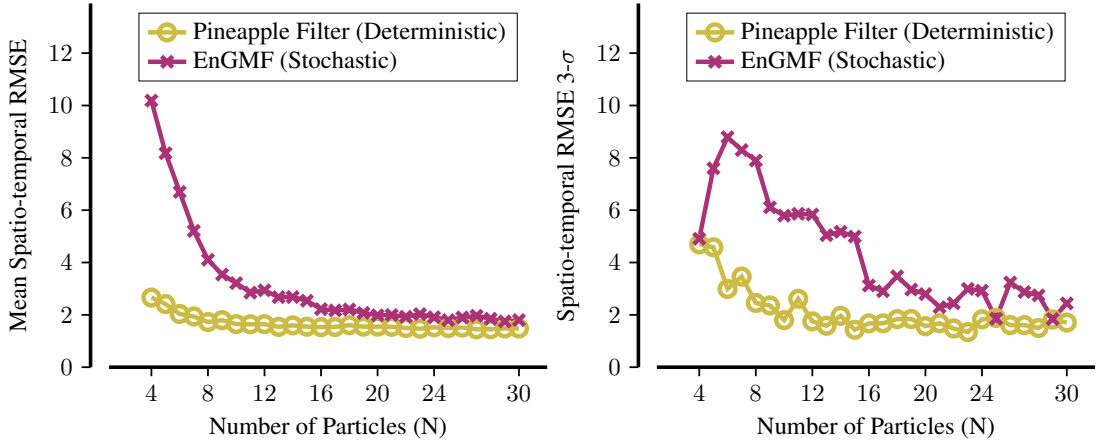

The Lorenz '63 equations are,
\begin{equation}\label{eq:lorenz63-equations}
\begin{aligned}
    x' &= \sigma(y - x),\\
    y' &= x(\rho - z),\\
    z' &= xy-\beta z,
\end{aligned}
\end{equation}
with standard parameters $\sigma = 10$, $\rho = 28$, and $\beta = \frac{8}{3}$.~\cite{lorenz1963deterministic}
The dynamics are propagated for a time interval of $0.12$ time units through an adaptive Runge-Kutta method. The dynamics are assumed to be exact, and no synthetic process noise is introduced.
The defensive factor~\cref{eq:defensive-factor} is set to $\mathfrak{v} = 0.1/\sqrt{N}$, ensuring convergence in the particle limit.

The non-linear measurement is the range from a fixed point of the system,
\begin{equation}
    h(x, y, z) = \sqrt{(x - c_x)^2 + (y - c_y)^2 + (z - c_z)^2},
\end{equation}
where,
\begin{equation}
    c_x = \sqrt{\beta(\rho-1)},\,c_y = \sqrt{\beta(\rho-1)},\,c_z = \rho-1,
\end{equation}
is the center of one of the wings of the butterfly in~\cref{fig:lorenz-63-dynamics}. The measurement variance is set to $R = 1$.

In order to account for errors, 192 (divisible by 12 to account for the amount of cores) Monte Carlo simulations of 220 steps were computed. The first 20 steps were discarded for all calculations to account for filter spinup.

For the error metric the spatio-temporal RMSE (see~\cite{popov2024adaptive}) was computed, and the mean was taken over all Monte-Carlo steps, with a 3-$\sigma$ bound computed. The results can be seen in~\cref{fig:lorenz63-experiment}.

As can be seen, the pineapple filter produces satisfactory results using $N=4$ particles for a $n=3$ dimensional system. The EnGMF on the other hand requires a minimum of $N=11$ particles for the same accuracy. The pineapple filter also produces estimates that track the truth with a significantly tighter error bound. The EnGMF suffers from spurious overconfidence in the error estimate, which the pineapple filter successfully avoids.

\subsection{NRHO example}

\begin{figure}[t]
    \centering
    \includegraphics[width=0.95\linewidth]{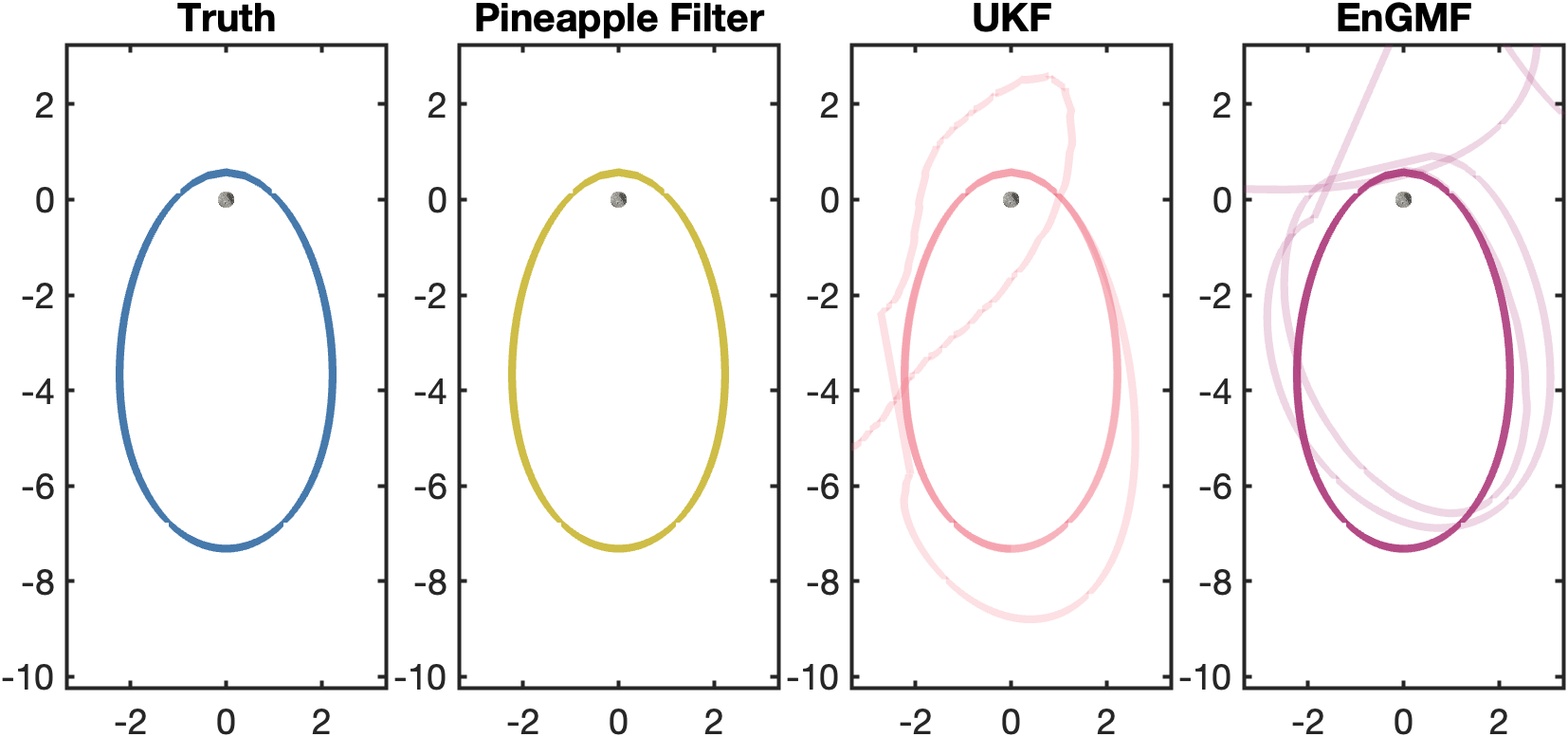}
    \caption{NRHO orbit MORE The scale on the $x$ and $y$ axis is in $10^7$ meters.}
    \label{fig:NRHO-orbit}
\end{figure}

Near-Rectilinear Halo Orbits (NRHOs) represent a class of stable cis-lunar orbits strategically positioned around the Moon~\cite{spreen2021trajectory}. 
In this work we model an NRHO with an orbital period of about $7.38$ days, using the Moon-centered Ephemeral $n$-body problem. 

A perfect circular orbit of the Moon around Earth is assumed. The gravitational pull of other objects such as the Sun is ignored. All other sources of perturbations such as solar wind are similarly ignored. A simple point-mass model of the moon's gravity is assumed.
The non-linear measurement is a range, range rate measurement, and two angles from the lunar north pole at around the time the space object should be at peri-lune, and thus at its highest velocity.

\begin{table}[htbp]
	\fontsize{10}{10}\selectfont
    \caption{Initial conditions for the ephimeral moon-centered three body problem}
   \label{tab:initial-conditions}
        \centering 
   \begin{tabular}{lr}
      $x$ & \SI{-1.203530188657220e10}{\meter}\\
      $y$ & \SI{1.114264238576134e10}{\meter}\\
      $z$ & \SI{9.905660048241707e+09}{\meter}\\
      $\dot{x}$ & \SI{-1.279778604446555e+04}{\meter/\second}\\
      $\dot{y}$ & \SI{1.036823000879196e+04}{\meter/\second}\\
      $\dot{z}$ & \SI{1.016485208254581e+04}{\meter/\second}
   \end{tabular}
\end{table}

The initial conditions of the space object are determined by the circular restricted three body problem, and are defined in~\cref{tab:initial-conditions}.
The initial conditions of the filters are perturbed by a covariance of,
\begin{equation}
    P_0 = \begin{bmatrix} 
    \SI{1e-2}{\meter^2} & & & & &\\
    & \SI{1e-2}{\meter^2} & & & &\\
    & & \SI{1e-2}{\meter^2} & & &\\
    & & & \SI{1e-8}{\meter^2/\second^2} & &\\
    & & & & \SI{1e-8}{\meter^2/\second^2} &\\
    & & & & & \SI{1e-8}{\meter^2/\second^2}
    \end{bmatrix},
\end{equation}
and the nonlinear measurement of range, range rate, and two angles,
\begin{equation}
\begin{gathered}
    h(x,y,z,\dot{x},\dot{y},\dot{z}) = \begin{bmatrix}
        \left\lVert p\right\rVert, & p\cdot v/\left\lVert p\right\rVert, & \tan^{-1}\left(y/x\right), & \tan^{-1}\left((z - r_{\mathleftmoon})/\left\lVert p\right\rVert\right)
    \end{bmatrix}^T\\
    p = \begin{bmatrix}x,& y,& z - r_{\mathleftmoon}\end{bmatrix}^T,\, v = \begin{bmatrix}\dot{x},& \dot{y},& \dot{z}\end{bmatrix}^T
    \end{gathered}
\end{equation}
where $r_{\mathleftmoon} = \SI{1740}{\kilo\meter}$ is the radius of the moon, and the measurement is perturbed by noise with a covariance of,
\begin{equation}
    R = \begin{bmatrix}
    \SI{1e2}{\meter^2} & & & \\
    & \SI{1e-2}{\meter^2/\second^2} & & \\
    & & \left(\ang{;;100}\right)^2  &\\
    & &  & \left(\ang{;;100}\right)^2
    \end{bmatrix},
\end{equation}
where $\ang{;;100}$ corresponds to 100 arcseconds.

The true space object that we are tracking performs station-keeping at apolune, by correcting the velocity to arrive at the same point relative to the moon for the next orbit. This is required as the model is unstable otherwise.
The filters account for station-keeping by assuming a process noise covariance of,
\begin{equation}
    Q = \frac{1}{4}P_0,
\end{equation}
which is the simplest way of tracking maneuvering space objects, and is known as `equivalent process noise'~\cite{zucchelli2024bayesian}.

The ensemble filters are run with $N=30$, and compared against the unscented Kalman filter (UKF) with default parameters $\alpha = 1$, $\beta = 2$, and $\kappa = -3$.~\cite{sarkka2023bayesian}
The defensive factor~\cref{eq:defensive-factor} is set to $\mathfrak{v} = 0.05$.
The filters are all run for 20 orbits and for 10 independent Monte Carlo runs.

The results of the experiment can be seen in~\cref{fig:NRHO-orbit}. As can be seen, both the UKF and the EnGMF fail to properly track the space object and fail spectacularly, predicting that the object is no longer in an orbit around the Moon. 
The Pineapple Filter (deterministic sampling) accurately tracks the object just from one measurement every orbit, which is likely impossible for a purely ``Gaussian'' (two-moment) filter.

\section{Conclusions}

This work presents a deterministic particle filter that has two surprising properties: \textit{i)} it requires very few particles for accurate estimation, and \textit{ii)} should converge to exact Bayesian inference (a proof of which is being explored in future research). This is the best of both worlds!
Other future  work will tackle the a deterministic optimal Gaussian sum update, and expand to non-Gaussian mixture model filters such as the ensemble Epanechnikov mixture filter~\cite{popov2024ensemble}.

\section{Acknowledgment}
This work is based on research that is in part sponsored by the Air Force Office of Scientific Research (AFOSR) under grant FA9550-23-1-0646. Part of this research was also funded by startup funds from the University of Hawai'i at M\=anoa.

\bibliographystyle{AAS_publication}   
\bibliography{bibfiles/monge,bibfiles/engmf,bibfiles/kernelapproximation,bibfiles/problems}   

\end{document}